\documentclass[aps,final,superscriptaddress,groupaddress,showkeys,floatfix,byrevtex,11pt]{revtex4-1}
\usepackage{graphicx}
\usepackage{dcolumn}
\usepackage{bm}
\usepackage[utf8x]{inputenc}
\usepackage{srctex}
\usepackage{amsmath}
\usepackage{amsfonts}
\usepackage{amssymb}
\usepackage{graphicx}
\usepackage{srctex}
\usepackage{xcolor}
\usepackage{subfigure}
\usepackage{cancel}

\newcommand{\ndpar}[3]{\frac{\partial^{#1}{#3}}{\partial {#2}^{#1}}}

\begin{document}

\title{Active Particles Moving in Two-Dimensional Space with Constant Speed: Revisiting the Telegrapher's Equation}

\author{Francisco J. Sevilla}
\email[]{fjsevilla@fisica.unam.mx}
\thanks{author to whom correspondence should be addressed.}
\affiliation{Instituto de F\'isica, Universidad Nacional Aut\'onoma de M\'exico, Apdo.\ Postal 20-364, 01000, M\'exico D.F., Mexico}

\author{Luis A. \surname{G\'omez Nava}}
\affiliation{Posgrado en Ciencias F\'isicas, Universidad Nacional Aut\'onoma de M\'exico}
\affiliation{Instituto de F\'isica, Universidad Nacional Aut\'onoma de M\'exico, Apdo.\ Postal 20-364, 01000, M\'exico D.F., Mexico}

\date{Today}

\begin{abstract}
Starting from a Langevin description of active particles that move with constant speed in infinite two-dimensional space and its corresponding Fokker-Planck equation, we develop a systematic method that allows us to obtain the coarse-grained probability density of finding a particle at a given location and at a given time to arbitrary short time regimes. By going beyond the diffusive limit, we derive a novel generalization of the telegrapher's equation. Such generalization preserves the hyperbolic structure of the equation and incorporates memory effects on the diffusive term.  While no difference is observed for the mean square displacement computed from the two-dimensional telegrapher's equation and from our generalization, the kurtosis results into a sensible parameter that discriminates between both approximations. We carried out a comparative analysis in Fourier space that shed light on why the telegrapher's equation is not an appropriate model to describe the propagation of particles with constant speed in dispersive media.
\end{abstract}


\maketitle

\section{Introduction}

The study of transport properties of active (self-propelled) particles has received much attention during the past two decades \cite{VicsekRev2012,RomanczukRev2012}. Self-propulsion, as a feature of systems out-of-equilibrium, has been introduced in a variety of contexts to describe, just to name a few, the foraging of organisms in ecology problems \cite{CodlingJRoySocInterface2008, ViswanathanBook}, the motion of bacteria \cite{CebersPRE2006}, and photon migration in multiple scattering media \cite{PolishchukPRE1996, RamakrishnaIJMPB2002, RamakrishnaPRE1999}. 

A simple model for self-propulsion is to consider that the particles move with constant speed in a manifold of interest, which in many cases coincides with the two-dimensional space.   This simplified modeling of particle activation has been approximately supported by experimental studies in many real biological systems \cite{BazaziCurrBio2008,BazaziProcRSocLondon2011,BodekerELett2010,EdwardsNature2007,GautraisJMathBio2009,LiPhysBiol2011} and has been used in several theoretical studies of systems that exhibit: collective motion \cite{VicsekPRL1995} for interacting self-driven particles, anomalous diffusion \cite{ChepizhkoPRL2013} when particles move in heterogeneous landscapes, or motion persistence \cite{WeberPRE2011, WeberPRE2012, RadtkePRE2012} if the particles are under the influence of fluctuating torques. 

Former studies on diffusion theory within the framework of random walks, used persistent random walks \cite[{}][{and references therein}]{WeissAdvChemPhys1883} and their phenomenological generalizations \cite{MasoliverPhysicaA1989} to incorporate internal states which sometimes are related to kinematic properties of the walker, such as velocity. Generally, the interest lies on a coarse-grained description of the probability density $P(\boldsymbol{x},t)$, of finding a particle at position $\boldsymbol{x}$ at time $t$, in which the detailed information about the internal states is irrelevant. As a standard procedure, the limit of the continuum is taken which leads to a partial differential equation for $P(\boldsymbol{x},t)$. Depending on the spatial dimension, those PDE are reminiscent of the well-known diffusion equation. For instance, the one-dimensional persistent random walk leads to the telegrapher's equation (TE) whose solution corrects, in the short time regime, the infinite speed of signal propagation exhibited by the solution to the diffusion equation. For longer times, the solution to the TE shows a transition, from a wave-like behavior at short times, to diffusion-like properties in the long-time regime \cite[{}][{and references therein}]{MasoliverEJP1996}. 

That dimensionality plays a significant role in various physical phenomena has been pointed out by many authors (see Ref. \cite{BarrowPhilTransRoySocLond1983} and references there in), particularly regarding the transmission of information described by the wave equation which favors three dimensions for signal fidelity transmission, a feature desired as a part of an anthropic principle. By contrast, the solution to the two-dimensional wave equation presents signal reverberation making impossible the transmission of sharply defined signals, additionally, the solution is negative for points inside the propagating front if initial data corresponds to an impulse with zero velocity \cite{MorseFeshbachBook}.  

Generalizations of the persistent random walk to arbitrary dimension $d$ greater than one have been formulated \cite{BogunaPRE1998}, however physical interpretation of the partial differential equation obtained after taking the limit of the continuum is hindered due to the presence of partial derivatives of order $2d$. This departure from the one-dimensional case, which contains at most partial derivatives of order two, is conspicuously important in the short time regime. Thus deriving an appropriate transport equation for the coarse-grained probability distribution in dimensions larger than one has been a central issue \cite{MasoliverPhysicaA1989,MasoliverPhysicaA1992,MasoliverPhysicaA1993,WeissPhysicaA2002}.

The description of particles that move with constant speed is also susceptible of the dimension of the system and one dimension seems to be particularly exceptional regarding the TE, since this last one has been derived exactly from various equivalent \emph{microscopic} models \cite{MasoliverPRE1994,SokolovPRE2003,KenkreSevilla2007} that consider the random transitions between the velocity states $\pm c$ (also known as Goldstein-Kac process, see Ref. \cite{PlyukhinPRE2010}) namely 
\begin{equation}
\ndpar{2}{t}{}P(x,t)+\gamma\ndpar{}{t}{}P(x,t)=c^{2}\ndpar{2}{x}{}P(x,t),
\label{Telegraphist_Eq}
\end{equation}
where $\gamma$ is the transition rate between states $\pm c$.
A generalization of this dichotomic process to dimensions larger than one \cite{RamakrishnaIJMPB2002} has lead also to a fourth-order PDE for particles that move with constant speed $\sqrt{2}c$ along the diagonals in two dimensions.

The straightforward generalization to $d\ge1$ spatial dimensions, $\partial^{2}_{t}P+\gamma\partial_{t}P=c^{2}\nabla_{d}P,$ has been considered before in the context of photon propagation in turbid media \cite{DurianJOSA1997, DurianOptLett1998, Ishimaru89p2210}, however, does not always result into an appropriate physical interpretation as has been already discussed in references \cite{MasoliverEJP1996,PorraPRE1997,GodoyPRE1997} particularly in two dimensions since at short times, the wave-like behavior, implies that the particle probability density becomes negative. 

In this work we present an analysis of Brownian-like particles that move with constant speed in infinite two-dimensional space and whose trajectories are obtained from Langevin-like equations. Through suitable transformations we are able to obtain approximated diffusion-like equations for the coarse-grained probability distribution $P(\boldsymbol{x},t)$. In the long-time limit we obtain the expected TE, and a novel generalization of it is obtained by going to a description of the system in a shorter-time regime. This generalization incorporates memory functions by keeping the hyperbolic nature of the original TE.

In section \ref{SectII} we provide the Langevin equations for the trajectories of particles that move with constant velocity and the Fokker-Planck equation for the probability density $P(\boldsymbol{x},\varphi,t)$ of a particle being at point $\boldsymbol{x}$, moving in the direction $\varphi$ at time $t$ is stated. In section \ref{SectIII} we present our method of analysis and derive a generalized TE. A comparative analysis between the generalized and the original TE is given in section \ref{SectIV}. We finally give our conclusion and final remarks in section \ref{SectV}.    

\section{\label{SectII}Langevin equations for Brownian agents with constant speed}
The kinematic state of a constant speed particle at time $t$ is determined by its position $\boldsymbol{x}(t)$ and the direction of motion $\hat{\boldsymbol{v}}(t)$, additionally, the particles are subject to the influence of stochastic fluctuations which only affect the direction of motion. The time evolution of the particle's state is given by the 
Langevin equations 
\begin{subequations}\label{modelo}
 \begin{align}
\frac{d{}}{dt}\boldsymbol{x}(t)&=v_{0}\, \hat{\boldsymbol{v}}(t),\\
\frac{d}{dt}\varphi(t)&=\xi(t),
 \end{align}
\end{subequations}
where the instantaneous unitary vector $\hat{\boldsymbol{v}}(t)$ is given by $(\cos\varphi(t),\sin\varphi(t))$, $\varphi(t)$ being the angle between the direction of motion and the horizontal axis. These equations describe the motion of a Brownian particle that moves with constant speed and changes its direction of motion due to Gaussian white noise $\xi(t)$, \emph{i.e.}, $\langle\xi\rangle=0$, $\langle\xi(t)\xi(s)\rangle=2\gamma\delta(t-s)$, where $\gamma$ is a constant that has units of $[time]^{-1}$ and denotes the intensity of the noise. Quantities with explicit time dependence denote those stochastic processes that appear in eqs. \eqref{modelo}, reserving the use of quantities without the explicit temporal dependence to appear in the corresponding Fokker-Planck equation.

From equations (\ref{modelo})  we obtain the following equation for 
the one particle probability density $P({\boldsymbol{x}},\varphi,t)\equiv \langle \delta({\boldsymbol{x}}-{\boldsymbol{x}}(t))\delta(\varphi-\varphi(t))\rangle$, 
\begin{equation}
\frac{\partial}{\partial t}P({\boldsymbol{x}},\varphi,t)+v_{0}\hat{\boldsymbol{v}}\cdot \nabla P({\boldsymbol{x}},\varphi,t)=-\frac{\partial}{\partial\varphi}
\langle \xi(t)\delta({\boldsymbol{x}}-{\boldsymbol{x}}(t))\delta(\varphi-\varphi(t))\rangle,
\end{equation}
where $\langle\cdot\rangle$ denotes the average over noise realizations. After making use of Novikov's theorem we get the Fokker-Planck equation
\begin{equation}
\frac{\partial}{\partial t}P({\boldsymbol{x}},\varphi,t)+v_{0}\, \hat{\boldsymbol{v}}\cdot \nabla P({\boldsymbol{x}},\varphi,t)=\gamma\frac{\partial^2 }{\partial \varphi^2}P({\boldsymbol{x}},\varphi,t).
\label{Fokker-Planck}
\end {equation}
In last expression we have omitted the term $-v_{0}\nabla\cdot\left\langle\left(\int_{0}^{t}ds\, \hat{\boldsymbol{v}}(s)\right)\delta({\boldsymbol{x}}-{\boldsymbol{x}}(t))\delta(\varphi-\varphi(t))\right\rangle$ assuming that the integral within parentheses vanishes at all times. Equation \eqref{Fokker-Planck} has also been derived from equivalent arguments in Ref. \cite{RamakrishnaPRE1999}.  

By performing the Fourier transform over the spatial coordinates and performing the Fourier expansion with respect to the angle $\varphi$, we transform equation (\ref{Fokker-Planck}) into the following set of tridiagonal coupled ordinary differential equations for the $n$-th coefficient of the expansion $\widetilde{P}_n({\boldsymbol{k}},t)$
\begin{widetext}
\begin{equation}
\frac{d}{dt}\widetilde{P}_n({\boldsymbol{k}},t)=-\frac{v_{0}}{2}\left[
\left(ik_{x}+	k_{y}\right)\widetilde{P}_{n-1}({\boldsymbol{k}},t)+
\left(ik_{x}-k_{y}\right)\widetilde{P}_{n+1}({\boldsymbol{k}},t)\right]-\gamma n^{2}\widetilde{P}_n({\boldsymbol{k}},t)
\label{ecuaciones_acopladas}
\end{equation}
\end{widetext}
that  satisfies $\widetilde{P}_n({\boldsymbol{k}},t)={\widetilde{P}}^{\ast}_{-n}(-\boldsymbol{k},t)$ and is given by $(2\pi)^{-2}\int d^{2}\boldsymbol{x}\int_{0}^{2\pi} d\varphi\, e^{i(\boldsymbol{k}\cdot\boldsymbol{x}-n\varphi)}\, P(\boldsymbol{x},\varphi,t).$  We are inte\-rested in the solution of (\ref{ecuaciones_acopladas}) with the initial condition: $\widetilde{P}_{n}(\boldsymbol{k},0)=\delta_{n,0}/2\pi$ which corresponds to the initial condition $P({\boldsymbol{x}},\varphi,0)=\delta^{(2)}({\boldsymbol{x}})/2\pi$,
where $\delta_{n,m}$, $\delta^{(2)}({\boldsymbol{x}})$ denote the Kronecker delta and the 2-dimensional Dirac delta res\-pectively. Through a further transformation, namely, 
\begin{equation}
 \widetilde{P}_n({\boldsymbol{k}},t)=e^{-\gamma n^2\,t}\widetilde{p}_n({\boldsymbol{k}},t),
\end{equation}
we obtain the one-step process with nonlinear coefficients
\begin{equation}
\frac{d}{dt}\widetilde{p}_n=-\frac{v_{0}}{2}\left[
\left(ik_{x}+	k_{y}\right)e^{-\gamma(-2n+1)\,t}\, \widetilde{p}_{n-1}+\right.
\left.\left(ik_{x}-k_{y}\right)e^{-\gamma(2n+1)\,t}\, \widetilde{p}_{n+1}\right],
\label{ecuaciones_acopladas2}
\end{equation}
where the arguments of $\widetilde{p}_{n}$ have been omitted for clarity. Although eqs. \eqref{ecuaciones_acopladas2} can be solved by the method of continued fractions \cite{Riskenbook}, we are interested in a coarse-grained description of the system in which the direction of motion of the particle is not relevant. Thus we focus on probability density distribution $P_{0}(\boldsymbol{x},t)\equiv(2\pi)^{-1}\int d^{2}\boldsymbol{k}\, e^{-i\boldsymbol{k}\cdot\boldsymbol{x}}\, \widetilde{p}_{0}(\boldsymbol{k},t)=\int_{0}^{2\pi}d\varphi\, P(\boldsymbol{x},\varphi,t)$.

The explicit appearance of the exponential factors in equations (\ref{ecuaciones_acopladas2}) makes clear that they are suitable to perform an analysis of different time regimes for $P_{0}(\boldsymbol{x},t),$ that initiate in the diffusive limit (long time regime) and extends to consider shorter times.

\section{\label{SectIII} The generalization of the telegrapher's equation}

Lets first consider the long time regime or diffusive limit, $3\gamma t\gg1$, for which we only hold the three Fourier coefficients $n=0,\, \pm1$ in eqs. (\ref{ecuaciones_acopladas2}), \emph{i.e.}
\begin{subequations}
 \begin{align}
  \frac{d}{dt}\widetilde{p}_{0}=&-\frac{v_{0}}{2}e^{-\gamma t}\left[\left(ik_{x}+k_{y}\right)\widetilde{p}_{-1}+\left(ik_{x}-k_{y}\right)\widetilde{p}_{1}\right]\\
  \frac{d}{dt}\widetilde{p}_{\pm1}=&-\frac{v_{0}}{2}e^{\gamma t}\left(ik_{x}\pm k_{y}\right)\widetilde{p}_{0}
 \end{align}
\end{subequations}
and by eliminating $p_{\pm1}$ one can show straightforwardly that $P_{0}(\boldsymbol{x},t)$ satisfies the TE
\begin{eqnarray}\label{2D_Telegrapher}
\frac{\partial^2}{\partial t^2}P_{0}(\boldsymbol{x},t)+\gamma\frac{\partial}{\partial t}P_{0}(\boldsymbol{x},t)=\frac{v_0^2}{2}\nabla^{2}P_{0}(\boldsymbol{x},t)
\end{eqnarray}
which agrees with the diffusive limit obtained in ref. \cite{PorraPRE1997} in the context of a transport equation that considers the scattering of the direction of motion.  This result is well known and corresponds to our first approximation to the problem. Equation (\ref{2D_Telegrapher}) describes wave-like propagation in the short time regime of pulses that travel not with speed $v_{0}$ but diminished by the factor $1/\sqrt{2},$ as is well known. In the asymptotic limit $\gamma t\rightarrow\infty$, the dispersive term dominates over the inertial one, given by the second order partial derivative with respect $t$, and equation (\ref{2D_Telegrapher}) reduces to the diffusion equation with diffusion constant $D=v_{0}^{2}/2\gamma$.  The solution to equation \eqref{2D_Telegrapher} is given explicitly in refs. \cite{MorseFeshbachBook,PorraPRE1997} and is solved under standard boundary conditions at infi\-nity, $P_{0}(\boldsymbol{x},t)\vert_{\vert\boldsymbol{x}\vert\rightarrow\infty}\rightarrow0,$ and under the initial conditions $P_{0}(\boldsymbol{x},0)=\delta^{(2)}(\boldsymbol{x})$, $\partial_{t}P_{0}(\boldsymbol{x},0)=0$, which are derived from the initial conditions for eqs. \eqref{ecuaciones_acopladas}, the last one, arises exactly from the coarsening procedure. In the wave-vector domain, the solution to \eqref{2D_Telegrapher} is simple and the same for arbitrary dimension $d$, given by $\widetilde{P}_{0}(\boldsymbol{k},0)\, e^{-\gamma t/2}\left[\gamma\sin\omega_{k{_d}}t /2\omega_{k{_d}}+\cos\omega_{k{_d}}t\right]$ with $\omega_{k{_d}}^{2}\equiv c_{d}^{2}k_{d}^{2}-\gamma^{2}/4,$ and $c_{d}\equiv v_{0}/\sqrt{d},$ $k_{d}$ the speed of propagation and the norm of the wave-vector in $d$ dimensions. At short times, the expression can be approximated by $\widetilde{P}_{0}(\boldsymbol{k},0)\cos c_{d}kt$, which corresponds to the normalized solution of the $d$-dimensional wave equation with initial conditions $P_{0}(\boldsymbol{k},0)$, $\partial_{t}P_{0}(\boldsymbol{k},0)=0$.

For a description of the system in a shorter time regime, namely $5\gamma t\gg1,$ we will require to take into account the next coefficients $n=\pm2$ of the Fourier expansion, thus from eq. \eqref{ecuaciones_acopladas2} we get
\begin{subequations}
 \begin{align}
  \frac{d}{dt}\widetilde{p}_{0}&=-\frac{v_{0}}{2}e^{-\gamma t}\left[\left(ik_{x}+k_{y}\right)\widetilde{p}_{-1}+\left(ik_{x}-k_{y}\right)\widetilde{p}_{1}\right]\label{p05modes}\\
  \frac{d}{dt}\widetilde{p}_{\pm1}&=-\frac{v_{0}}{2}e^{\gamma t}\left[\left(ik_{x}\pm k_{y}\right)\widetilde{p}_{0}+e^{-4\gamma t}\left(ik_{x}\mp k_{y}\right)\widetilde{p}_{\pm2}\right]\\
  \frac{d}{dt}\widetilde{p}_{\pm2}&=-\frac{v_{0}}{2}e^{3\gamma t}\left(ik_{x}\pm k_{y}\right)\widetilde{p}_{\pm1}.
 \end{align}
\end{subequations}
These equations lead to a novel generalization of the TE for $P_{0}(\boldsymbol{x},t)$ after eliminating $\widetilde{p}_{\pm1}$ from \eqref{p05modes}, namely
\begin{equation}\label{GTE}
\frac{\partial^2}{\partial t^2}P_{0}(\boldsymbol{x},t)+\gamma\frac{\partial}{\partial t}P_{0}(\boldsymbol{x},t)=v_{0}^{2}\,\nabla^{2}
\times\int_{0}^{t}ds\, \phi(t-s)P_{0}(\boldsymbol{x},s)+\frac{v_{0}^{2}}{4}e^{-4\gamma t}Q(\boldsymbol{x})
\end{equation}
where the memory function $\phi(t)$ that appears in last expression is given by $\frac{3}{4}\delta(t)-\gamma e^{-4\gamma t}$ and $Q(\boldsymbol{x})$ is a function that is determined from the initial distribution $P(\boldsymbol{x},\varphi,0)$ through 
\begin{equation}
\int_{0}^{2\pi}d\varphi\, \left[e^{i2\varphi}\left(\partial_{x}+i\partial_{y}\right)^{2}+e^{-i2\varphi}\left(\partial_{x}-i\partial_{y}\right)^{2}-\right.
\left.\left(\partial^{2}_{x}+\partial^{2}_{y}\right)\right]P(\boldsymbol{x},\varphi,0). 
\label{Q}
\end{equation}
Boundary and initial conditions for (\ref{GTE}) can be shown to be the same as in the previous approximation.  

In the time regime of validity of equation (\ref{GTE})we have the following solution in the Fourier-Laplace domain
\begin{equation}\label{GTE_solutionFourierLaplace}
\widehat{P}_{0}(\boldsymbol{k},\epsilon)=\frac{(\epsilon +\gamma)\widetilde{P}_{0}(\boldsymbol{k},0)+(v_{0}^{2}/4)\widetilde{Q}(\boldsymbol{k})/(\epsilon+4\gamma)}{\epsilon^{2}+\gamma\epsilon + v_{0}^{2}k^{2}\hat{\phi}(\epsilon) }
\end{equation}
where $\hat{\phi}(\epsilon)=3/4-\gamma(4\gamma+\epsilon)^{-1}$, $\epsilon$ denotes the Laplace variable and $\hat{f}(\epsilon)=\int_{0}^{\infty}dt\, e^{-\epsilon t}f(t)$ is the Laplace transform of $f(t).$ Inversion of the green function $G(\boldsymbol{k},\epsilon)=[\epsilon^{2}+\gamma\epsilon + v_{0}^{2}k^{2}\hat{\phi}(\epsilon)]^{-1}$ can be done approximately  
in the time and space regimes, $\epsilon\ll4\gamma,$ $k\ll4\gamma/v_{0}$ respectively (see the appendix B) giving for $P_{0}(\boldsymbol{x},t)$ 
\begin{equation}\label{GTE_solution}
G(\boldsymbol{x},t)=\frac{8\gamma}{\pi v_{0}^{2}}\int d^{2}\boldsymbol{x}^{\prime}\, e^{-\frac{8\gamma}{v_{0}^{2}t}(\boldsymbol{x}-\boldsymbol{x}^{\prime})^{2}}G_{TE}(\boldsymbol{x}^{\prime},t)
\end{equation}
where $G_{TE}(\boldsymbol{x},t)$ is the corresponding well known Green's function of the TE \cite{MorseFeshbachBook,PorraPRE1997}.

As shown in the following section, equation \eqref{GTE} gives an appropriate description of Brownian particles that move with constant speed, however, by considering the following coefficients of the Fourier expansion, $n=\pm3,$ partial derivatives of order four start to appear and the memory function turns more involved.

\section{\label{SectIV} Discussion}

It has been established that the TE gives a better description of particles that move with constant speed than the diffusion equation \cite{PorraPRE1997}. Thus, how better is the generalization of the former, equation \eqref{GTE}, at shorter times? To answer this question we calculate the mean square displacement (MSD) and the kurtosis $\kappa$ for the solutions of equations \eqref{2D_Telegrapher} and \eqref{GTE}, and we compare them with the exact results from numerical simulations by solving equations \eqref{modelo}.

\begin{figure}
 \includegraphics[width=0.7\textwidth]{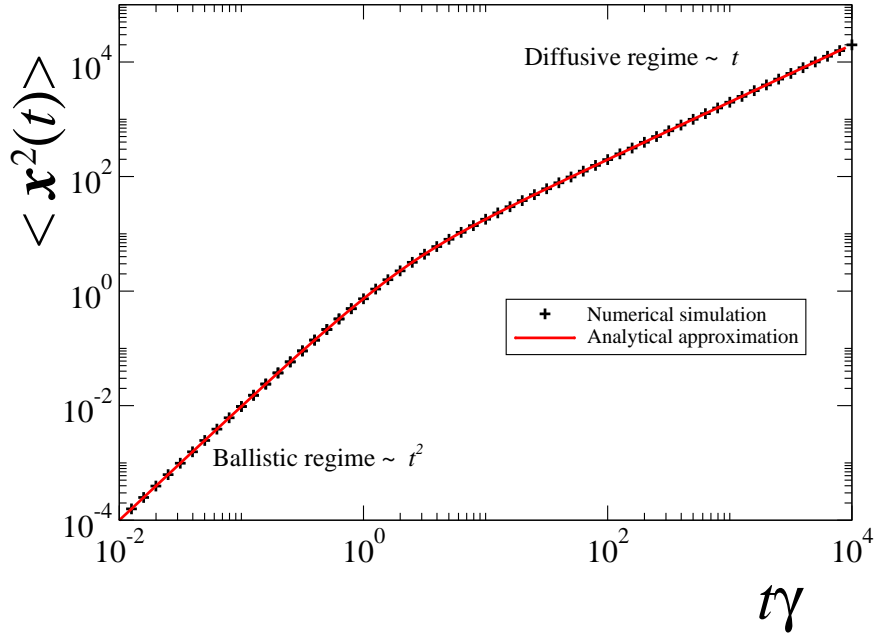}
 \caption{(Color online) Mean square displacement (MSD) in units of $\left(v_{0}/\gamma\right)^{2}$ \emph{vs.} $\gamma t.$ The continuous (red) line is the analytical approximation given by the expression (\ref{MSD_GTE}), the data showed in symbols were obtained by averaging the numerical solution of equations (\ref{modelo}) considering $10^{5}$ trajectories and integrating over 2$\times 10^{6}$ time-steps.}
 \label{Figure_MSD}
\end{figure}

\paragraph{The Mean Square Displacement} A prediction from our analysis is the mean squared displacement: $\langle \boldsymbol{x}^2(t)\rangle=\int d^{2}\boldsymbol{x}\, \boldsymbol{x}^{2}P_{0}(\boldsymbol{x},t)=-\left(\partial_{kx}^{2}+\partial_{ky}^{2}\right)\widetilde{P}_{0}({\boldsymbol{k}},t)\left.\right\vert_{\boldsymbol{k}=0}$. By multiplying by $\boldsymbol{x}^{2}$ an integrating over the whole space equations (\ref{2D_Telegrapher}) and (\ref{GTE}), we obtain respectively
\begin{subequations}\label{MSD_eq}
\begin{align}
\frac{d^{2}}{dt^{2}}\langle \boldsymbol{x}^2
(t)\rangle+\gamma \frac{d}{dt}\langle \boldsymbol{x}^2(t)\rangle&=2v_{0}^{2}\\
\frac{d^{2}}{dt^{2}}\langle \boldsymbol{x}^2(t)\rangle+\gamma \frac{d}{dt}\langle \boldsymbol{x}^2(t)\rangle&=2v_0^2+v_0^2e^{-4\gamma t}\left(1+\beta/4\right)\label{msd2}
\end{align}
\end{subequations}
where $\beta=\int d^{2}\boldsymbol{x}\, \boldsymbol{x}^{2}Q(\boldsymbol{x}).$ It can be shown that $\beta=-4$ for all smooth circularly symmetric initial distributions (see Appendix). This observation assures that the MSD for both approximations show the same and exact dependence with time, namely
\begin{align}
\langle \boldsymbol{x}^2(t)\rangle=&\langle \boldsymbol{x}^{2}(0)\rangle+\frac{4D}{\gamma}\left[\gamma t -\left(1 - e^{-\gamma t}\right)\right],
\label{MSD_GTE}
\end{align}
for circularly-symmetric distributions taken as initial distributions. Thus, the MSD does not provide a measure of the departure between the solution of equation (\ref{2D_Telegrapher}) and (\ref{GTE}), neither between these and the exact solution obtained from numerical simulations (see fig. \ref{Figure_MSD}). Equations \eqref{MSD_eq} were solved with the initial condition $\frac{d}{dt}\langle \boldsymbol{x}^{2}(t)\rangle\vert_{t=0}=0,$ which has been chosen based on the expected physical behavior. For $\gamma t\ll1$ the particles show the $t^{2}$ dependence for short times that corresponds to the ballistic regime.  

The linear dependence for long times is evident and leads to the effective diffusion constant $D\equiv v_{0}^{2}/2\gamma$. 
Expression (\ref{MSD_GTE}) coincides with the result for the normal Brownian motion, \emph{i.e.} with fluctuating speeds, in two dimensions if in the expression for the diffusion constant, $v_{0}$ is substituted by $\sqrt{\langle\boldsymbol{v}^{2}\rangle}=\sqrt{2k_{B}T/m}$, where $k_{B}$ the Boltzmann constant, $T$ the absolute temperature and $m$ the mass of the particles.

\paragraph{Kurtosis} A sensible parameter to measure the departure between our results given by the equation (\ref{GTE}), the TE (\ref{2D_Telegrapher}), and the exact result from numerical simulations, is given by the kurtosis $\kappa$ which has been used as a measure to test multivariate normal distributions \cite{Mardia74p115}, is given  explicitly by 
\begin{equation}\label{kurtosisdef}
\kappa = \left\langle\left[(\boldsymbol{x}-\langle\boldsymbol{x}\rangle)^{T}\Sigma^{-1}(\boldsymbol{x}-\langle\boldsymbol{x}\rangle)\right]^{2}\right\rangle,
\end{equation}
where $\boldsymbol{x}^{T}$ denotes the transpose of the vector $\boldsymbol{x}$ and $\Sigma$ is the matrix defined  by the average of the dyadic product $(\boldsymbol{x}-\langle\boldsymbol{x}\rangle)^{T} \cdot(\boldsymbol{x}-\langle\boldsymbol{x}\rangle)$

For circular symmetric distributions it reduces to  
\begin{equation}\label{kurtosisradial}
\kappa=4\frac{\langle\vert\boldsymbol{x}\vert^{4}\rangle_{rad}}{\langle\vert\boldsymbol{x}\vert^{2}\rangle_{rad}^{2}} 
\end{equation}
where $\langle\cdot\rangle_{rad}$ denotes the average over the radial distribution $rP(r)$, \emph{i.e.} $\int_{0}^{\infty}dr\, r \, P(r)\, (\cdot). $
For two-dimensional Gaussian distributions $\kappa$ takes the invariant value, \emph{i.e.} independent of width and mean, 8 and it can be shown following the lines in Appendix \ref{AppKurtosis}, from the two-dimensional wave equation, that the circularly-symmetric, normalized, solutions with zero initial velocity has a kurtosis value 8/3. 
  
We have from equations (\ref{2D_Telegrapher}) and (\ref{GTE}) that the kurtosis in each case, is given by (see Appendix \ref{AppKurtosis}) 
\begin{subequations}\label{kurtosis}
\begin{align}
 \kappa_{3}&=8 \left[\gamma^{2}t^2-2 \gamma t\left(2+ e^{-\gamma t}\right)+6\left(1-e^{-\gamma t}\right) \right]
 \left[\gamma t-\left(1-e^{-\gamma t}\right)\right]^{-2},\label{k3}\\
 \kappa_{5}&=8 \left[\gamma^{2}t^2- 5\, \gamma t\left(\frac{3}{4}+\frac{1}{3} e^{-\gamma t}\right)+\left(\frac{87}{16}-\frac{49}{9}e^{-\gamma t}\right.\right.
 \left.\left.+\frac{1}{12^{2}}e^{-4\gamma t}\right) \right]\left[\gamma t-\left(1-e^{-\gamma t}\right)\right]^{-2}\label{k5},
\end{align}
\end{subequations}
where the subindex denotes the number of Fourier modes retained.
 
\begin{figure}
 \includegraphics[width=0.7\textwidth]{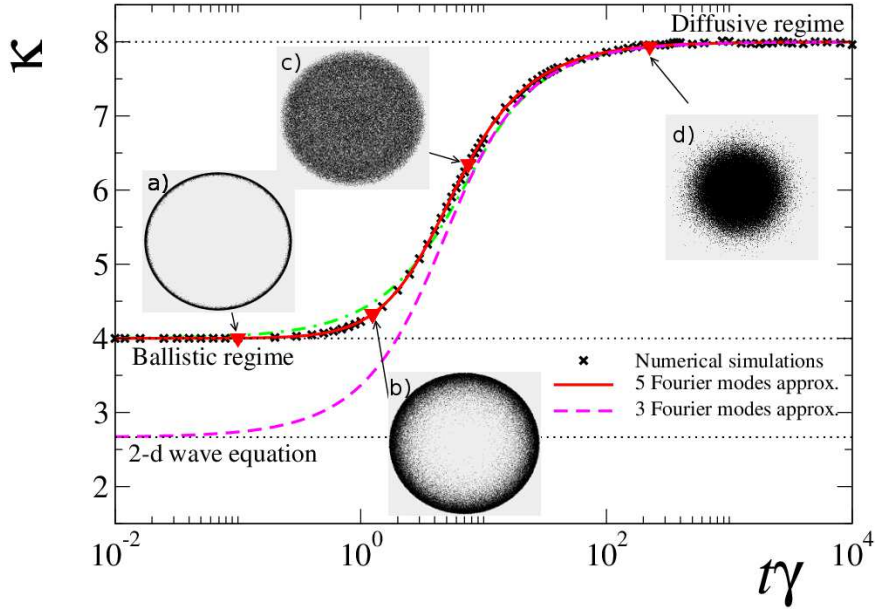}
 \caption{(Color online) Kurtosis $\kappa$ for the circularly-symmetric solutions of the TE \eqref{2D_Telegrapher} (dashed-magenta line), eq. \eqref{GTE} (continuous-red line) and of the exact solution obtained from numerical simulations of eqs. \eqref{modelo} (cross symbols) \emph{vs} $t\gamma$. Doted lines mark the values 8, 4 and $8/3\simeq2.6667$ that correspond to the values of $\kappa$ for: the two-dimensional Gaussian distribution, the two-dimensional distribution of particles that move with constant speed, the circularly-symmetric solutions of the two-dimensional wave equation, respectively. The insets show snapshots obtained from numerical simulations of the particles distribution $P_{0}(\boldsymbol{x},t)$ when they start to move from the origin at four different values of $t\gamma,$ whose values of $\kappa$ are characteristics (solid-red triangles): a) $t\gamma=0.1$, $\kappa=4.003$; b) $t\gamma=1.25$, $\kappa=4.33$; c) $t\gamma=7.5$, $\kappa=6.352$ and d) $t\gamma=225$, $\kappa=7.938$ (the radius of the distribution has been scaled to fit the viewing area). Numerical simulations were performed by averaging $10^{5}$ trajectories computed from eqs. (\ref{modelo}) and integrating over 2$\times 10^{6}$ time-steps. The dotted-dashed-green line shows the time dependence of the kurtosis of the symmetric solution of the non-homogeneous TE obtained in ref. \cite{MasoliverPhysicaA1993}.}
 \label{Figure_kurt}
\end{figure}

These results are compared with the exact calculation of $\kappa$ in fig. \ref{Figure_kurt}. As expected, both descriptions and exact numerical calculations lead asymptotically to a Gaussian distribution since $\kappa\rightarrow8$ as $\gamma t\rightarrow\infty$. On the other hand, in the short time limit $\gamma t\ll1$ a discrepancy between both descriptions is conspicuous. From the TE (\ref{2D_Telegrapher}) the kurtosis of the distribution goes to $8/3$ which coincides with the value for the two-dimensional wave equation (see dashed-magenta line in fig. \ref{Figure_kurt}). However, from our numerical calculations, the time dependence of the kurtosis of the distribution of Brownian particles that move with constant speed acquire the value $4$ for shot times (see cross symbols in fig. \ref{Figure_kurt}), and coincides with the kurtosis for the distribution that solves the generalized TE \eqref{GTE} at all times.  

The inset a) in fig. \ref{Figure_kurt} shows a propagating ring-like distribution of particles at $t\gamma=0.1$ for which $\kappa=4.003$. $\kappa$ rises as the ring starts to being fill, as can be appreciated in inset b), for which $t\gamma=1.25$ and $\kappa=4.33$. In the inset c), the ring is full and the distribution is approximately homogeneous on the disk. This is reflected in the value $\kappa=6.352$ at $t\gamma=7.5$ ($\kappa=16/3\simeq5.333$ for a uniform distribution on a disk of given radius). At longer times the distribution becomes Gaussian as indicated by the value $\kappa=7.938$, as shown in inset d) for $t\gamma=225$.

It is worth to point out that though the kurtosis calculated from the rotationally-symmetric solutions of eq. \eqref{GTE} coincides with the exact result computed from the Langevin eq. \eqref{modelo}, it does not show the characteristic hollow inside the ring in the short time regime shown in inset a) of fig. \ref{Figure_kurt}). In fact the solutions of related two-dimensional telegrapher-like equations, as the one presented in this work (another is presented in ref. \cite{MasoliverPhysicaA1993} for the two-dimensional persistent random walk) show up a \emph{wake} effect that is characteristic of the solution of the two-dimensional wave equation \cite{MorseFeshbachBook}. 

We point out that although the solution of the inhomogeneous TE, obtained in ref. \cite{MasoliverPhysicaA1993}, does give the values $\kappa=4$ and 8 at short and long times respectively, it differs from our results in the intermediate regime (see dotted-dashed-green line in figure \ref{Figure_kurt}). 

\begin{figure}[h]
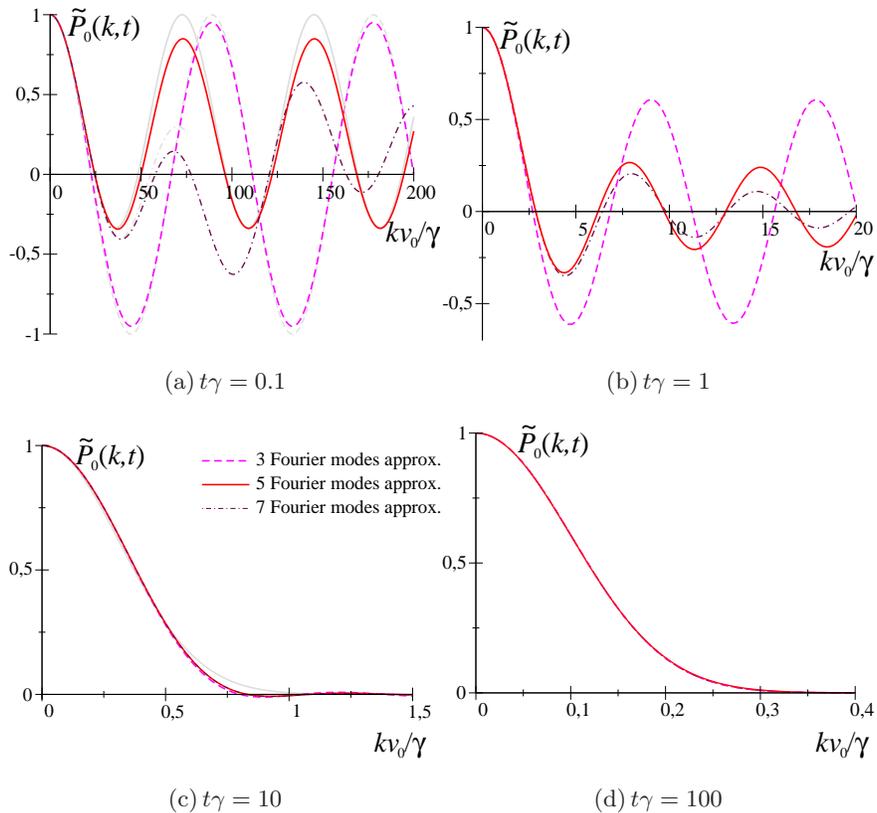

\subfigure[\,$t\gamma=0.1$]{\includegraphics[width=0.35\textwidth]{Fig3a}}\subfigure[\,$t\gamma=1$]{\includegraphics[width=0.35\textwidth]{Fig3b}}\\
\subfigure[\,$t\gamma=10$]{\includegraphics[width=0.35\textwidth]{Fig3c}}\subfigure[\,$t\gamma=100$]{\includegraphics[width=0.35\textwidth]{Fig3d}}
\caption{(Color online) Plots showing the probability density $\widetilde{P}_{0}(k,t)$, as function of $k=\vert\boldsymbol{k}\vert$ at four different times, namely: $\gamma t=0.1$, $\gamma t=1.0$, $\gamma t=10.0$, and $\gamma t=100.0$ in panels (a), (b), (c) and (d), respectively. Continuous-red line corresponds to the solution of eq. \eqref{GTE} while the dashed-magenta line to the solution of the TE \eqref{2D_Telegrapher}. The numerical solution, considering up to seven Fourier modes in eqs. \eqref{ecuaciones_acopladas2}, is shown by the dashed-doted-maroon line.  The dashed-gray line in panel (a) corresponds to the normalized solution to the two-dimensional wave equation with propagation speed $v_{0}\sqrt{2}$, namely $\cos v_{0}kt/\sqrt{2}$ while the continuous-gray one in the same panel, corresponds to the short time limit of the Laplace inversion of expression \eqref{GTE_solutionFourierLaplace} given by eq. \eqref{shortime_GTEsol} in the Appendix \ref{AppGTE}.}
\label{Fourier_graphics}
\end{figure}

We finish this section by comparing the probability density distributions obtained in Fourier space in the short and long time regimes. Due to the simple appearance of the Laplace operator in eqs. \eqref{2D_Telegrapher}, \eqref{GTE} and the symmetrical initial conditions, their respective solutions in Fourier space depends simply on $k=\vert\boldsymbol{k}\vert.$ In the short time regime, $\gamma t=0.1$ panel (a) in fig. \ref{Fourier_graphics}, the solution to the TE (dashed-magenta line) has a close behavior to the normalized solution to the 2-dimensional wave equation (dashed-gray line). Thus, those features attached to the solution of the two-dimensional wave equation are inherited by the two-dimensional TE. On the other hand the solution to eq. \eqref{GTE} (solid-red line) departs conspicuously from that wave-like behavior improving, as shown by the previous results, the description of particles that move with constant speed. In this regime, the equation satisfies the inhomogeneous wave equation $\partial^{2}_{tt}P_{0}(\boldsymbol{x},t)=(3/4)v_{0}^{2}\nabla^{2}P_{0}(\boldsymbol{x},t)+(v_{0}^{2}/4)Q(\boldsymbol{x})$ (see Appendix \ref{AppGTE}), whose solution is given by \eqref{shortime_GTEsol} and shown by the continuous-gray line in panel (a) of fig. \ref{Fourier_graphics}. We also show $\widetilde{P}_{0}$ obtained from the numerical solution of eqs. \eqref{ecuaciones_acopladas2} considering the first seven Fourier modes (dashed-dotted-maroon line). It is evident from panels (a) and (b) that more than five modes are needed to describe accurately the time evolution of $P_{0}(\boldsymbol{x},t)$ and we point out that $J_{0}(kv_{0}t)$, which corresponds to the two-dimensional Fourier transform of $\delta(\vert\boldsymbol{x}\vert-v_{0}t)/2\pi\vert\boldsymbol{x}\vert$, approximates well the seven modes solution up to $kv_{0}t\simeq50.$    

At $\gamma t=1$ (panel (b) in fig. \ref{Fourier_graphics}) the solution to eq. \eqref{2D_Telegrapher} departs from its wave-like behavior and starts to resemble the solution of eq. \eqref{GTE}, while the latter is qualitatively similar for the one provided by the seven modes approximation.  For longer times $\gamma t=10,\, 100$, panels (c) and (d) respectively, the three approximations shown are closer to each other and tend to a Gaussian (continuous-gray line that is solution of the diffusion equation with diffusion constant $v_{0}^{2}/2\gamma$). 

\section{\label{SectV} Conclusions and Final Remarks}
Starting from a Langevin formalism to describe active particles that move with constant speed, $v_{0},$ in two dimensions. We obtained a Fokker-Planck for $P(\boldsymbol{x},\varphi,t)$, the probability density of finding a particle at position $\boldsymbol{x}$ moving in the direction $(\cos\varphi,\sin\varphi)$ at time $t.$ By using Fourier transforms we obtained an infinite system of coupled ordinary differential equations for the Fourier modes $\widetilde{P}_{n}(\boldsymbol{k},t)$. By a suitable transformation we were able to do a systematic analysis for different time regimes that tends towards a shorter time description of the coarse-grained probability $P_{0}(\boldsymbol{x},t).$ Our formalism allows us, in principle, to obtain solutions arbitrarily close to the exact one by taking into account higher Fourier modes.

The long time or diffusive approximation, considers only the first three Fourier modes and lead to the well known TE \eqref{2D_Telegrapher} with a propagation speed $v_{0}/\sqrt{2}$. A shorter time description that takes into account the next two Fourier modes, lead to a generalized TE. Such equation is inhomogeneous and the generalization relies in the non-Markovian nature of the diffusive term. A comparison between both approximations was made by computing the second and fourth moments of the circularly-symmetric solutions of both equations, namely the mean square displacement and the kurtosis. The former did not exhibit a difference between the two descriptions, while the latter, being a measure for the shape of the probability density, resulted into a sensible parameter in the short time regime. However, despite the outstanding agreement between the numerical results and the kurtosis given by our generalization of the TE, the latter could not describe a shorter time regime were sharply signals are transmitted as shown in the inset a) of fig. \ref{Figure_kurt}.

From our analysis it is clear why the TE is not an appropriate model to describe the dynamics of Brownian particles that move with constant speed in the short time regime, namely, only the three lowest modes of the Fourier expansion of the joint probability density $P(\boldsymbol{x},\varphi,t)$ are considered, however, the drift term $v_{0}\hat{\boldsymbol{v}}\cdot\nabla P(\boldsymbol{x},\varphi,t)$ in \eqref{Fokker-Planck} induce important correlations among the rest of the Fourier modes. Those correlations are damped as the fluctuating direction of motion $(\cos\varphi(t),\sin\varphi(t))$ becomes Gaussian distributed. The numerical calculation considering up to the first seven Fourier shows that in the short time regime, there exist strong correlations between the particle position and its direction of motion.    

We comment, in passing, that the appearance of me\-mory in equation (\ref{GTE}), makes it suitable to consider it as a candidate to describe anomalous diffusion phenomena as does a former generalization of the TE that use fractional-time derivatives \cite{CompteJPhysA1997}. Indeed, if for the moment we disregard the non-homogeneous term, the mean square displacement for arbitrary memory function $\phi(t)$ is
\begin{equation}
 \langle\boldsymbol{x}^{2}(t)\rangle=4v_{0}^{2}\int_{0}^{t}ds\, e^{-\gamma(t-s)}\int_{0}^{s}ds^{\prime}\int_{0}^{s^{\prime}}ds^{\prime\prime}\phi(s^{\prime\prime}).
\end{equation}
If the memory function $\phi(t)$ decays algebraically for large times as $t^{-\alpha},$ with $0<\alpha<1$, as does the Mittag-Leffler function $\tau^{-1}E_{\alpha,1}\left(-t^{\alpha}/\tau^{\alpha}\right)$, $\tau$ a constant with units of time and $E_{\mu,\nu}(z)=\sum_{n=1}^{\infty}z^{n}/\Gamma(\mu j+\nu)$, the mean square displacement can be expressed as
\begin{equation}
 \langle\boldsymbol{x}^{2}(t)\rangle=4\frac{v_{0}^{2}}{\tau}\int_{0}^{t}ds\, e^{-\gamma(t-s)} s^{2}E_{\alpha,3}\left(-s^{\alpha}/\tau^{\alpha}\right)   
\end{equation}
which behaves super-diffusively, $\sim t^{2-\alpha},$ in the asymptotic limit.

Extensions of our analysis would consider colored noise and/or the effects of interactions among particles.

\begin{acknowledgments}
 FJS acknowledges support from DGAPA-UNAM through the grant PAPIIT-IN113114.
\end{acknowledgments}

\appendix
\section{\label{AppQ}The second and fourth moments of the circularly-symmetric inhomogeneous term $Q(\boldsymbol{x})$}
If smooth circularly-symmetric initial distributions of the form $P(\boldsymbol{x},\varphi,0)=\mathcal{X}(x)/(2\pi)^{2}$ are considered, with $x=\vert\boldsymbol{x}\vert$, expression \eqref{Q} reduces simply to 
\begin{equation}
Q(\boldsymbol{x})=-\frac{1}{2\pi}\left[\frac{1}{x}\frac{\partial}{\partial x}\left(x\frac{\partial}{\partial x}\right)\right]\mathcal{X}(x)
\end{equation}
since the terms proportional to $\int_{0}^{2\pi}d\varphi\, e^{\pm i2\varphi}$ are zero. Thus, the factor  $\beta=\int d^{2}\boldsymbol{x}\, \boldsymbol{x}^{2}Q(\boldsymbol{x})$ that appears in \eqref{msd2} is obtained by computing
\begin{equation*}
\beta=-\int_0^{\infty}dx\,  x^2\left[\frac{\partial \mathcal{X}(x)}{\partial x}+x\frac{\partial^2 \mathcal{X}(x)}{\partial x^2}\right]
\end{equation*}
where the Laplacian in polar coordinates has been used. Integrating by parts once the second term in brackets and using boundary conditions $\mathcal{X}(x)\left\vert_{x=\infty}\right.=0,
\frac{\partial \mathcal{X}(x)}{\partial x}\left\vert_{x=\infty}\right.=0$ we get that
\begin{equation*}
\beta=2\int_0^{\infty} x^2\frac{\partial \mathcal{X}(x)}{\partial x}dx.
\end{equation*}
Integrating by parts again and using that $\int dx\, x\mathcal{X}(x)=1$ we finally arrive to the result
\begin{align*}
\beta=-4.
\end{align*}

Analogously, the fourth moment is given by 
\begin{equation*}
\int_0^{\infty}dx\,  x^4\left[\frac{\partial \mathcal{X}(x)}{\partial x}+x\frac{\partial^2 \mathcal{X}(x)}{\partial x^2}\right]=2^{4}\int_0^{\infty}dx\,  x^3\mathcal{X}(x)
\end{equation*}
which gives zero for the localized initial condition $\mathcal{X}(x)=\delta(x)/x.$

\section{\label{AppKurtosis}Kurtosis of the solutions of eqs. \eqref{2D_Telegrapher} and \eqref{GTE}}

Expressions \eqref{kurtosis} are obtained as follows. Eqs. \eqref{2D_Telegrapher} and \eqref{GTE} are multiplied by $x^{4}\, xdx$ (recalling $x=\vert\boldsymbol{x}\vert$) and are integrated, over $x$, from 0 to $\infty$. For circularly-symmetric solutions we get
%
%
\begin{subequations}
\begin{equation}
\frac{d^{2}}{dt^{2}}\langle \vert\boldsymbol{x}\vert^4
(t)\rangle_{rad}+\gamma \frac{d}{dt}\langle \vert\boldsymbol{x}\vert^4
(t)\rangle_{rad}=
8v_{0}^{2}\langle \vert\boldsymbol{x}\vert^2
(t)\rangle_{rad}
\end{equation}
\begin{equation}
\frac{d^{2}}{dt^{2}}\langle \vert\boldsymbol{x}\vert^4
(t)\rangle_{rad}+\gamma \frac{d}{dt}\langle \vert\boldsymbol{x}\vert^4
(t)\rangle_{rad}=
4^{2}v_0^2\int_{0}^{t}ds\, \phi(t-s)\langle \vert\boldsymbol{x}\vert^2
(s)\rangle_{rad}
\end{equation}
\end{subequations}
As shown in Appendix \ref{AppQ} the inhomogeneous term of eq. \eqref{GTE} does not contribute for the initial condition in which all particles depart from the origin. The solutions to the last equations, for vanishing initial conditions, are
\begin{subequations}
\begin{equation}
\langle \vert\boldsymbol{x}\vert^4(t)\rangle_{rad}=
8v_{0}^{2}\int_{0}^{t}ds\, e^{-\gamma(t-s)}\int_{0}^{s}ds^{\prime}\langle \vert\boldsymbol{x}\vert^2
(s^{\prime})\rangle_{rad},
\end{equation}
\begin{equation}
\langle \vert\boldsymbol{x}\vert^4(t)\rangle_{rad}=4^{2}v_0^2\int_{0}^{t}ds\, e^{-\gamma(t-s)}\times\\
\int_{0}^{s}ds^{\prime}\int_{0}^{s^{\prime}}ds^{\prime\prime}\phi(s^{\prime}-s^{\prime\prime}) \langle \vert\boldsymbol{x}\vert^2
(s^{\prime\prime})\rangle_{rad},
\end{equation}
\end{subequations}
respectively. After substitution of the MSD \eqref{MSD_GTE} in the last equations and performing the integration we get, for the kurtosis given by \eqref{kurtosisradial}, expressions \eqref{kurtosis}. 

\section{\label{AppGTE}Approximate solution of the eq. (\ref{GTE})}

The solution in Fourier-Laplace domain to the GTE given by expression (\ref{GTE_solutionFourierLaplace}) can be computed by inverting the Green function
\begin{equation}\label{Green}
\widehat{G}({k},\epsilon)=\left[{\epsilon^{2}+\gamma\epsilon + v_{0}^{2}k^{2}\left(3/4-\gamma\left[\epsilon+4\gamma\right]^{-1}\right)}\right]^{-1}\\
\xrightarrow[\epsilon\ll4\gamma]{} \left[\epsilon^{2}+\gamma_{k}\epsilon+\frac{v_{0}^{2}k^{2}}{2}\right]^{-1}
\end{equation}
where $\gamma_{k}\equiv \left(1+\frac{v_{0}^{2}k^{2}}{2^{4}\gamma^{2}}\right)\gamma$ and we have explicitly used that $\hat{\phi}(\epsilon)=3/4-\gamma(\epsilon+4\gamma)^{-1}.$ By defining the $k$-dependent frequency $\omega_{k}^{2}\equiv\frac{v_{0}^{2}k^{2}}{2}-\left(\frac{\gamma_{k}}{2}\right)^{2}$ the Laplace inversion of the left hand side can be carried out  and (\ref{Green}) is given by
\begin{equation}\label{GreenApprox}
\widetilde{G}({k},t)\approx\frac{e^{-\gamma_{k}t/2}}{\omega_{k}}\sin\omega_{k}t
\end{equation}
for $4\gamma t\gg1$. This approximated Green function generalizes the corresponding one of the TE which is obtained by putting $k=0$ in $\gamma_{k}$, namely $\widetilde{G}^{0}({k},t)=e^{-\gamma t /2}\sin\omega_{0k}/\sin\omega_{0k}$ with $\gamma=\gamma_{k=0}$ and $\omega_{0k}=\left(v_{0}^{2}k^{2}/2\right)-\left(\gamma/2\right)^{2}.$

By using (\ref{GreenApprox}) we get
\begin{widetext}
\begin{multline}
 \widetilde{P}_{0}(\boldsymbol{k},t)\approx\frac{e^{-\gamma_{k}t/2}}{\omega_{k}}\left[\omega_{k}\cos\omega_{k}t+\left(\gamma-\frac{\gamma_{k}}{2}\right)\sin\omega_{k}t\right]\widetilde{P}_{0}(\boldsymbol{k},0)+\\
 \frac{v_{0}^{2}}{4}\widetilde{Q}(\boldsymbol{k})\frac{e^{-\gamma_{k}t/2}}{\omega_{k}}\frac{e^{-(4\gamma-\gamma_{k}/2)t}\omega_{k}-\omega_{k}\cos\omega_{k}t+\left(4\gamma-\gamma_{k}/2\right)\sin\omega_{k}t}{\left(\gamma_{k}/2-4\gamma\right)^{2}+\omega_{k}^{2}}
\end{multline}
\end{widetext}
Analytical inversion of the Fourier transform seems to be intractable by the appearance of $k^{4}$ in $\omega_{k}$, however for large space ranges, we have that to first order in $v_{0}^{2}k^{2}/2^{4}\gamma^{2}\ll1$, $\omega_{k}^{2}\approx\left(v_{0}^{2}k^{2}/2\right)\left(1-2^{-4}\right)-\left(\gamma/2\right)^{2}$ which resembles its corresponding counterpart of the TE. With these considerations the Green function can be approximated further by $e^{-v_{0}^{2}k^{2}t/2^{4}\gamma}\widetilde{G}^{0}({k}^{\prime},t)$ where $k^{\prime}=\sqrt{15}k/2^{2}.$ 

Thus the Green function of the GTE in coordinate space can be written as
\begin{equation}
G(\boldsymbol{x},t)=\frac{1}{4\pi D^{\prime}t}\int d\boldsymbol{x}^{\prime}\, e^{-(\boldsymbol{x}-\boldsymbol{x}^{\prime})/4D^{\prime}t} G^{0}(\boldsymbol{x}^{\prime},t)
\end{equation}
where $D^{\prime}=v_{0}^{2}/2^{4}\gamma.$

On the other hand, in the short time regime, expression \eqref{GTE_solutionFourierLaplace} can be approximately written as 
\begin{equation}
 \widehat{P}_{0}(\boldsymbol{k},\epsilon)=\frac{\epsilon\, \widetilde{P}_{0}(\boldsymbol{k},0)}{\epsilon^{2}+\frac{3}{4}v_{0}^{2}k^{2}}+\frac{v_{0}^{2}}{4}\frac{\widetilde{Q}(\boldsymbol{k})}{\epsilon^{2}+\frac{3}{4}v_{0}^{2}k^{2}}
\end{equation}
which after inverting the Laplace transform we obtain
\begin{equation}\label{shortime_GTEsol}
 \widetilde{P}_{0}(\boldsymbol{k},t)=\widetilde{P}_{0}(\boldsymbol{k},0)\left(\cos\frac{\sqrt{3}}{2}v_{0}kt+\frac{2}{3}\sin^{2}\frac{\sqrt{3}}{4}v_{0}kt\right),
\end{equation}
where we have imposed the rotational symmetry to write $\widetilde{Q}(\boldsymbol{k})=k^{2}\widetilde{P}_{0}(\boldsymbol{k},t)$. The term $\cos\frac{\sqrt{3}}{2}v_{0}kt$ in the last expression is reminiscent of the solution of the wave equation with a speed of propagation, $\sqrt{3}v_{0}/4,$ larger than the value $v_{0}/\sqrt{2}$ given by the TE. It can be checked in a direct manner that expression \eqref{shortime_GTEsol} satisfies the inhomogeneous wave equation
\begin{equation}
 \frac{\partial^{2}}{\partial t^{2}}P_{0}(\boldsymbol{x},t)=\frac{3}{4}v_{0}^{2}\nabla^{2}P_{0}(\boldsymbol{x},t)+\frac{v_{0}^{2}}{4}Q(\boldsymbol{x}).
\end{equation}


\end{document}